# Addressing unobserved heterogeneity at road user level for the analysis of conflict risk at tunnel toll plaza: A correlated grouped random parameters logit approach with heterogeneity in means

**Penglin Song**
Department of Civil and Environmental Engineering
The Hong Kong Polytechnic University
Hung Hom, Kowloon, Hong Kong
Email: *peng-lin.song@connect.polyu.hk*

**N.N. Sze (Corresponding author)**
Department of Civil and Environmental Engineering
The Hong Kong Polytechnic University
Hung Hom, Kowloon, Hong Kong
Tel: +852 2766-6062; Email: *tony.nn.sze@polyu.edu.hk*

**Ou Zheng**
Department of Civil, Environmental and Construction Engineering
University of Central Florida
Orlando, FL 32816-2450, United States
Email: *ouzheng1993@knights.ucf.edu*

**Mohamed Abdel-Aty**
Department of Civil, Environmental and Construction Engineering
University of Central Florida
Orlando, FL 32816-2450, United States
Email: *M.Aty@ucf.edu*

# ABSTRACT

Toll plaza is a designated area of controlled-access roads like expressway, bridge, and tunnel for toll collection. A number of toll booths are often placed at the toll plaza accommodating high passing traffic and multiple payment methods. Traffic and safety characteristics of toll plazas are different from that of other road entities. Different conflict risk indicators, which are usually longitudinal, have been adopted for real-time safety assessment. In this study, correlated grouped random parameter logit models with heterogeneity in the means are established to capture the unobserved heterogeneity, with additional flexibility, at road user level for the association between conflict risk and influencing factors. In addition, modified conflict risk indicator is developed to assess the safety of diverging, merging, and weaving movements of traffic, with which vehicles' dimensions (width and length), and longitudinal and angular movements are considered. Also, prevalence and severity of both rear-end and sideswipe conflicts are assessed. Results indicate that toll collection type, vehicle's location, average longitudinal speed, angular speed, acceleration, and vehicle class all affect the risk of traffic conflicts. Furthermore, there are significant correlation among the random parameters of severe traffic conflicts. Proposed analytic method can accommodate the conflict risk analysis for different conflict types and account for the correlation of unobserved heterogeneity. Findings should shed light on appropriate remedial measures like traffic signs, road markings, and advanced traffic management system that can improve the safety at tunnel toll plazas.

# 1. INTRODUCTION

Toll plaza is a designated area for toll collection of controlled-access roads like expressways, bridges, and tunnels. Traffic and safety characteristics of toll plazas are different from that of other road entities because of the differences in geometric design, traffic management and control (*Wong et al., 2006*). More importantly, weaving, diverging, and merging movements of traffic approaching the toll booths, especially when multiple toll collection (i.e., manual, and electronic) methods are available. Manual toll payment vehicles need to decelerate when approaching the toll booths, while electronic toll payment vehicles can continue to travel through the toll plaza at a relatively high speed. Derivation of the speed of mixed traffic can increase the crash risk of toll plaza (*Abdelwahab and Abdel-Aty, 2002*). It is crucial to identify the factors that affect the safety of toll plaza, hence effective countermeasures can be developed. Studies have assessed the safety of toll plaza, merging, and diverging areas, based on historical crash data. For example, toll plaza layout, horizontal curves, toll collection method, and traffic signs and road markings are found associated with the crash risk at toll plazas (*Wong et al., 2006*; *Sze et al., 2008*; *Abuzwidah et al., 2014*; *Abuzwidah and Abdel-Aty, 2015*, *2018*). In addition, lighting condition is associated with the occurrence and severity level of crash at diverging areas (*Mergia et al., 2013*). Furthermore, geometric design characteristics including number of lanes, road alignment, and length of deceleration lane are associated with the crash occurrence at off-ramp areas (*Chen et al., 2009*; *Chen et al., 2011*; *Calvi et al., 2012*). Last but not least, association between crash occurrence and possible risk factors can be moderated by collision type (i.e., rear-end, sideswipe, and angle collisions) (*Guo et al., 2019*).

Road safety analysis based on historical crash data is often subject to the problems like under-reporting, misclassification, and imbalanced crash data (*Tsui et al., 2009*; *Lord and Mannering, 2010*; *Chen et al., 2022*; *Ding et al., 2022*). To this end, it is possible to estimate the safety risk based on real-time traffic data collected using video observational survey and driving simulator approaches (*Sayed et al., 2013*; *Yun et al., 2017*; *Chen et al., 2019*; *Saad et al., 2019*; *Arun et* al., 2021a; Li et al., 2022; Wang et al., 2022). Surrogate safety measures like time and distance headway, mean and deviation of speed, acceleration rate, and traffic conflicts can be applied to assess the safety level of road entities. To estimate the risk of traffic conflicts, indicators like time-to-collision, post-encroachment time, and deceleration rate to avoid the crash are used (Tarko, 2018). According to the "safety pyramid", traffic incidents can be classified into three categories: (i) normal interactions; (ii) traffic conflicts; and (iii) crashes (Hydén, 1987). As a crash is the extreme form of traffic conflict, modeling the latter (which requires shorter observation period to accrue enough sample) may provide a reliable foundation for better understanding of crash mechanisms (El-Basyouny and Sayed, 2013; Sayed et al., 2013; Zheng et al., 2021). Despite that, more work is required for accurate prediction of crash severity based on traffic conflict analysis (Arun et al., 2021; Paul and Ghosh, 2021).

In most of the previous studies, the risk of traffic conflicts has been estimated based on vehicle length, position of centroid, and longitudinal movement of conflicting vehicles only (Sacchi and Sayed, 2016; Wang et al., 2019; Xu et al., 2021; Yang et al., 2021). It may result in underestimation of traffic conflict risk and bias in parameter estimates when vehicle width, and motions in two dimensions of conflicting vehicles are not considered. This is particularly true for the interactions between vehicles at intersections, and diverging, merging, and weaving areas. To this end, it is necessary to consider vehicle width and length (Arun et al., 2021b; Arun et al., 2021c), two-dimensional (i.e., longitudinal and angular) movements (Ward et al., 2015; Tarko, 2021), and point of contact (Laureshyn et al., 2010; Jiménez et al., 2013) of conflicting vehicles in traffic conflict analysis.

For the analytic methods, discrete outcome models have been applied to model the association between risk of conflicts and possible factors like road geometry, environmental condition, traffic flow, and driver characteristics (Sacchi and Sayed, 2016; Uzondu et al., 2018). However, there are unobserved and unmeasurable factors that may affect the association between observed variables of interest and conflict risk. For example, there are significant differences in (unobserved) safety perception, attitude, and behavior among the drivers who are of the same age. Hence, driver age may not be able to fully account for the effect of unobserved individual heterogeneity on driver performance and associated conflict risk (*Guo et al., 2018*; *Zhu et al., 2022*). To this end, the random parameter approach can be applied to model the risk of traffic conflicts, accounting for the effect of unobserved heterogeneity (*Chen et al., 2021*). Several crash frequency and severity studies have considered heterogeneity in the means and variances of the random parameters to capture unobserved heterogeneity with additional flexibility (*Behnood and Mannering, 2017a, 2017b*; *Seraneeprakarn et al., 2017*; *Xin et al., 2017*; *Alnawmasi and Mannering, 2019*; *Behnood and Mannering, 2019*; *Al-Bdairi et al., 2020*; *Huo et al., 2020*; *Islam et al., 2020*; *Islam and Mannering, 2020*; *Yu et al., 2020*; *Hou et al., 2021*; *Li et al., 2021*; *Se et al., 2021*; *Song et al., 2021*; *Yan et al., 2021a, 2021b*; *Zamani et al., 2021*; *Alnawmasi and Mannering, 2022a, 2022b*; *Alogaili and Mannering, 2022*; *Wang et al., 2022a*; *Wang et al., 2022b*). Furthermore, to account for correlation among random parameters and unobserved effects due to repeated observations of the same entities, correlated grouped random parameters model can be adopted for crash severity analysis (*Mannering et al., 2016*; *Fountas et al., 2018a*; *Fountas et al., 2018b*; *Saeed et al., 2019*; *Hou et al., 2020*; *Fountas et al., 2021*; *Meng et al., 2021*). However, the aforementioned studies were based on historical crash record, not many studies have considered the heterogeneity in the means of the random parameters and correlation among random parameters in real-time conflict risk estimation at the micro-level (*Zhang et al., 2021*).

In recent years, it is increasingly popular to collect traffic data using unmanned aerial vehicles (also known as drones) for traffic and safety analysis (*Stipancic et al., 2016*; *Wang et al., 2019*). Based on the aerial footage captured by the drone, it is possible to extract high-resolution vehicle trajectory data using automated image recognition and processing technique (*Krajewski et al., 2018*; *Li et al., 2020*). Although several studies have assessed the safety risk of toll plazas based on drone data, they do not explicitly account for the effects of vehicle dimensions and conflict type on risk estimation (*Xing et al., 2019*; *Xing et al., 2020a*; *Xing et* *al., 2020b*). For example, it is possible to distinguish among various conflict types, i.e., head-on, sideswipe, and angled conflicts using pixel-based image classification technique based on high resolution vehicle trajectory data (*Wu et al., 2020*).

In this study, modified traffic conflict indicator, taking into account vehicle length and width, angular and longitudinal movements, and conflict type (i.e., rear-end and sideswipe), is proposed to assess the safety risk at a tunnel toll plaza, based on high-resolution vehicle trajectory data obtained from drone video. Then, the correlated grouped random parameter multinomial logit approach with heterogeneity in the means of the random parameters is adopted to measure the association between conflict risk at tunnel toll plaza and possible factors, including vehicle class, speed and acceleration of vehicle, toll collection type, and spatial characteristics, for which effects of unobserved heterogeneity and correlation among random parameters at the road user level are accounted for.

The remainder of this paper is organized as follows. Data collection, model formulation, and analysis method are described in **Section 2**. Then, **Section 3** summarizes the data used. Furthermore, estimation results and interpretations are presented in **Section 4**. Finally, concluding remarks and suggestions for future research would be given.

## 2. METHOD

### 2.1 Traffic conflict

Time-to-collision refers to the time required for two conflicting vehicles to collide if their speed and path remain unchanged (*Hayward, 1972*). For the rear-end collision of two vehicles travelling in the same direction, time-to-collision (TTC) can be calculated as,

$$TTC = \begin{cases} \dfrac{x_l - x_f - L_l}{v_f - v_l}, & if\ v_f > v_l \\ \infty, & if\ v_f > v_l \end{cases} \quad (1)$$

where $x_l$ is the displacement of front bumper of leading vehicle, $x_f$ is that of following vehicle, $v_l$ is the speed of leading vehicle, $v_f$ is that of following vehicle, and $L_l$ is the length of leading vehicle.

However, Equation (1) may not be capable of modeling the risk of angle and sideswipe collisions for diverging, merging, and weaving traffic. To this end, dimensions and angular movement of conflicting vehicles should be considered. **Figure 1** illustrates the typical interaction between two conflicting vehicles at the toll plaza, and diverging and merging areas. As shown in **Figure 1**, paths of Vehicle 1 and Vehicle 2 are intersecting at angle $\alpha$. In addition, rectangle *1A1B1C1D* and *2A2B2C2D* denote the areas covered by Vehicle 1 and Vehicle 2, respectively. Shaped area covered by parallelogram *abcd* represents the overlapping area of trajectories of Vehicle 1 and 2 if their paths remain unchanged.

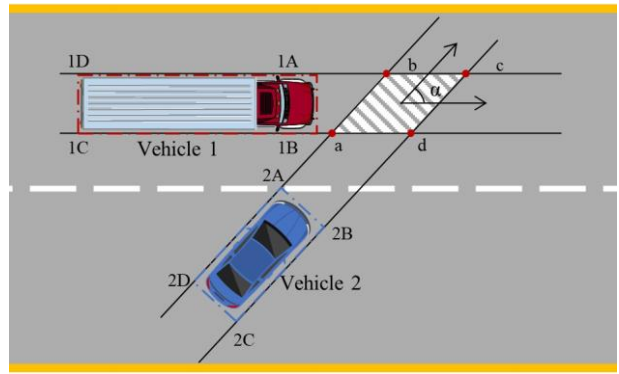

**Figure 1 Illustration of interaction between two conflicting vehicles**

Lengths, widths, and movement directions of Vehicles 1 and 2 can affect the location, shape, and size of the overlapping area. Point of contact for potential collision can be predicted, based on the assumption that Vehicles 1 and 2 would collide if their path and speed remain unchanged. Let $t_{pq}$ denote the time at which the corner ($p$) of a vehicle reaches that ($q$) of the shaped area, where $p$ is *1A*, *1B*, *1C*, …, and *2D*, and $q$ is *a*, *b*, *c*, and *d*, respectively. There may be a collision when the front of one vehicle reaches the overlapping area before another vehicle completely leaves the area. For example, when $t_{1Ba} < t_{2Aa}$ and $t_{1Ca} > t_{2Aa}$, the front left corner (*2A*) of Vehicle 2 will hit the right side (*1B1C*) of Vehicle 1 at *a* (Interested reader is referred to **Appendix** for all possible collision scenarios). Hence, time-to-collision can be estimated based on the difference in arrival time at the overlapping area between the conflicting vehicles, with which the two-dimensional vehicle motion is considered. One should note that a leading vehicle refers to the vehicle that arrives at the overlapping area first, based on the instantaneous motion of conflicting vehicles at the time of observation, in the subsequent analysis. Time step interval for the analysis depends on the frame rate of video (*Laureshyn et al., 2010*; *Gu et al., 2019*).

In general, collision can be classified into four categories: (i) head-on collision, (ii) angle collision, (iii) sideswipe collision, and (iv) rear-end collision, based on the point of contact and intersecting angle of conflicting vehicles (*Wu et al., 2020*). In this study, maximum intersecting angle of the sample is less than nine degrees. Hence, only the sideswipe and rear-end conflicts are considered.

**Figure 2** illustrates some possible conflict scenarios in this study. For example, rear-end conflict refers to that when the front of a vehicle hit the rear of another vehicle, and sideswipe conflict refers to that when the corner of a vehicle hit the side of another vehicle.

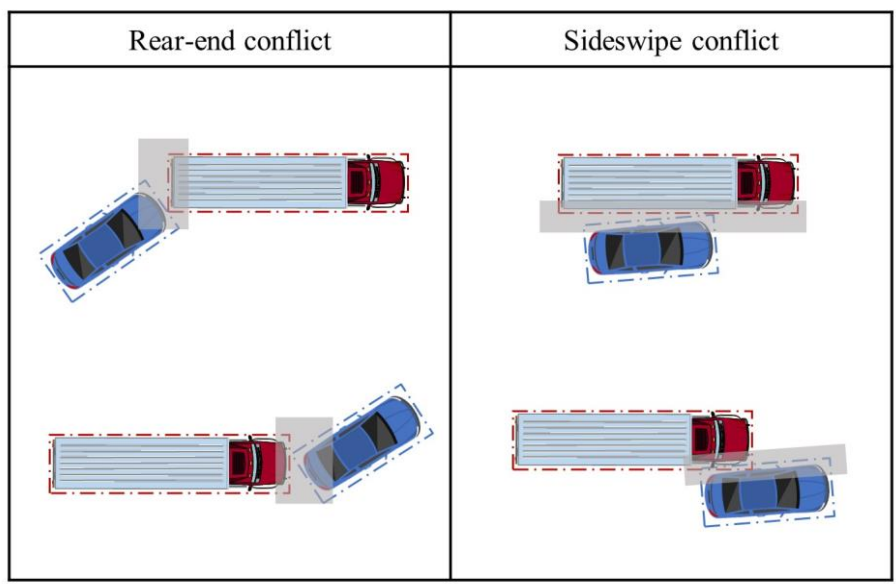

**Figure 2 Illustration of possible conflict scenarios**

In conventional traffic conflict analyses, minimum time-to-collision (*Johnsson et al., 2021*) and instantaneous time-to-collision (*Gu et al., 2019*; *Xing et al., 2020a*) are commonly used to predict the occurrence of a traffic conflict. In this study, instantaneous time-to-collision is adopted as the conflict indicator, accounting for possible endogeneity of speed-related variables. When the value of instantaneous time-to-collision is lower than a pre-determined threshold, a traffic conflict will exist. In general, threshold of time-to-collision ranges from one to three seconds (*Madsen and Lahrmann, 2017*; *Essa and Sayed, 2019*; *Essa et al., 2019*; *Wang et al., 2021*), and more than one stratification points can be established to distinguish among conflicts of different severity levels (*Essa et al., 2019*). In this study, two commonly used thresholds of time-to-collision are adopted: (i) 3 seconds for the occurrence of slight conflicts; and (ii) 1.5 seconds for the occurrence of severe conflicts. It should be noted that "conflict severity" refers to how close it is a collision may occur. It does not necessary imply the occurrence of severe crash. Therefore, it is more likely for a severe conflict to become a crash, compared to slight conflict (*Hydén, 1987*). Furthermore, a "traffic conflict" will be defined only when there is no traffic congestion (*Johnsson et al., 2021*), and the overlapping area is located within the pre-defined study area.

## 2.2 Model formulation

In this study, risk of traffic conflict at the toll plaza is modeled as discrete outcomes, namely (i) no conflict (i.e., normal interaction), (ii) slight conflict, and (iii) severe conflict. Hence, multinomial logit regression approach is adopted to measure the association between conflict risk and possible influencing factors. For instance, probability of interaction $i$ that has outcome $j$ ($j \in J$) is given by:

$$P_{ij} = P(U_{ij} > U_{ik}), \qquad \forall k \neq j \tag{2}$$

where $U_{ij}$ is the function that determines the probability of outcome $j$ for interaction $i$ that is given by:

$$U_{ij} = \boldsymbol{\beta}_j' \boldsymbol{x}_i + \varepsilon_{ij}, i = 1, \ldots \ldots, n \tag{3}$$

where $\boldsymbol{x}_i$ is a vector of explanatory variables for interaction $i$, $\boldsymbol{\beta}_j$ is a vector of mean coefficients for outcome $j$, $\varepsilon_{ij}$ is the error term which is assumed to be independent and identically distributed (IID) with Type 1 extreme value (Gumbel) distribution, and $n$ is the total number of observations.

Then, the probability that outcome $j$ will occur for interaction $i$ can be expressed as:

$$P_{ij}|\beta_j = \frac{\exp(\boldsymbol{\beta}_j' \boldsymbol{x}_i)}{\sum_{j=1}^{J} \exp(\boldsymbol{\beta}_j' \boldsymbol{x}_i)} \tag{4}$$

And, the unconditional probability can be computed as:

$$P_{ij} = \int \frac{exp(\boldsymbol{\beta}_j' \boldsymbol{x}_i)}{\sum_{j=1}^{J} exp(\boldsymbol{\beta}_j' \boldsymbol{x}_i)} f(\boldsymbol{\beta}|\boldsymbol{\varphi}) \, d\boldsymbol{\beta} \tag{5}$$

where $f(\boldsymbol{\beta}|\boldsymbol{\varphi})$ is the density function for vector $\boldsymbol{\beta}$, and $\boldsymbol{\varphi}$ is the vector of parameters that defines the density function.

In addition, the correlated random parameters approach with heterogeneity in the means is applied, accounting for the effect of unobserved heterogeneity. Hence, the coefficients would be modified as:

$$\boldsymbol{\beta}_{jr} = \boldsymbol{\beta}_j + \boldsymbol{\Theta}_j z_j + \boldsymbol{\Gamma} \boldsymbol{\omega}_j \tag{6}$$

where $\boldsymbol{\beta}_{jr}$ is a vector of random parameters, $\boldsymbol{\Theta}_j$ is a matrix of estimated parameters, $z_j$ is a vector of explanatory variables that capture heterogeneity in the means, $\boldsymbol{\omega}_j$ is normally distributed with $N(0, \sigma_j^2)$, and $\boldsymbol{\Gamma}$ is the Cholesky matrix based on Cholesky decomposition. Considering the effect of repeated observations from the same entities (i.e., vehicle interactions), grouped parameters are incorporated into the model. Parameters are allowed to vary across groups of observation with:

$$\boldsymbol{\beta}_g = \boldsymbol{\beta} + \boldsymbol{\Gamma} \boldsymbol{\omega}_g \tag{7}$$

where $g$ denotes the group.

$\boldsymbol{\Gamma}$ matrix is a lower triangular matrix given as:

$$\boldsymbol{\Gamma} = \begin{bmatrix} \gamma_{1,1} & & & & \\ \gamma_{2,1} & \gamma_{2,2} & & & \\ \vdots & \vdots & \ddots & & \\ \gamma_{k-1,1} & \gamma_{k-1,2} & \cdots & \gamma_{k-1,k-1} & \\ \gamma_{k,1} & \gamma_{k,2} & \cdots & \gamma_{k,k-1} & \gamma_{k,k} \end{bmatrix} \quad (8)$$

The off-diagonal elements of $\boldsymbol{\Gamma}$ are usually set to be zero in conventional random parameter models. Thus, no correlation among random parameters is implied. To capture the possible correlations among random parameters, the off-diagonal elements should be non-zeros. One should note that the Cholesky matrix can be constrained (partially set to zero) (***Hensher et al., 2015***). The variance-covariance matrix can be written as:

$$\boldsymbol{C} = \boldsymbol{\Gamma}\boldsymbol{\Gamma}^T \quad (9)$$

which the diagonal elements are the standard deviations of random parameters, and off-diagonal elements are the covariance between random parameters, respectively. Standard deviation of the correlated random parameters can be expressed as:

$$\sigma_j = \sqrt{\gamma_{k,1}^2 + \gamma_{k,2}^2 + \cdots + \gamma_{k,k-1}^2 + \gamma_{k,k}^2} \quad (10)$$

$t$-statistics is used to assess the statistical significance of standard deviations of the correlated grouped random parameters. Standard error of the standard deviation is given by:

$$SE = \frac{S}{\sqrt{N}} \quad (11)$$

$$t = \frac{\sigma_j}{SE} \quad (12)$$

where $S$ is the standard deviation, and $N$ is the number of observations.

The correlation coefficient between two random parameters is computed as,

$$Cor(x_p, x_q) = \frac{cov(x_p, x_q)}{\sigma_p \sigma_q} \quad (13)$$

where $cov(x_p, x_q)$ is the covariance between the random parameters of variable $x_p$ and $x_q$, and $\sigma_p$ and $\sigma_q$ are the standard deviations of the random parameters.

Likelihood-ratio test is used to assess the goodness-of-fit of two competing models with the chi-square test statistics given by,

$$X^2 = -2[LL(\beta_1) - LL(\beta_2)] \tag{14}$$

where $LL(\beta_1)$ and $LL(\beta_2)$ are the log-likelihood functions at convergence of Model 1 and Model 2, respectively, and degree of freedom is equal to the differences in number of parameters between competing models.

Parameters can be estimated by simulated maximum likelihood method (*Train, 2009*; *Chen et al., 2020*; *Zhu and Sze, 2021*; *Zhu et al., 2021*). Previous studies indicate that 1,000 Halton draws is sufficient for the convergence of parameter estimation (*Meng et al., 2021*; *Yan et al., 2022*).

## 3. DATA

Toll plaza (Kowloon bound) of Cross-Harbour Tunnel in Hong Kong (left hand driving rule applies) is selected as the study site. Cross-Harbour Tunnel, which was opened in 1972, is the busiest among the three underwater crossings connecting Kowloon and Hong Kong Island. In 2019, annual average daily traffic of Cross-Harbour Tunnel was 106,679 (*Hong Kong Transport Department, 2020*). As shown in **Figure 3**, number of lanes increase from three (near the tunnel portal) to eight (near the toll booths) when travelling along the toll plaza. Of the eight toll booths, three are allocated for electronic toll collection [i.e., Lane 1 (bus-only lane), Lane 2, and Lane 8], and five are for manual toll collection. Speed limit going through the electronic toll booths is 50 kph. As the driver behaviors and associated safety risks may change when drivers are approaching the toll booths from different distances, the study area is stratified into three (i.e., Zone 1, Zone 2, and Zone 3). In Zone 2, the number of lanes starts to increase, lane changing activities are frequent. In Zone 3, lane changing activities are partially restrained, particularly for electronic toll collection lanes.

To capture the aerial video, drone (DJI Mavic Air 2) is used in this study. Height of the drone is 100 meters above ground, and the field of view is 84 degrees. Observation survey was conducted during the daytime on 8 weekdays in October of 2020. Weather was fine (i.e., sunny and no wind) in the observation period. Also, there was no traffic jam. Overall, 120-minute video was captured. Resolution of the video was 1080p and frame rate was 30 fps. Vehicle trajectories were extracted from the video using the Automated Roadway Conflicts Identify System (ARCIS) of the University of Central Florida's Smart and Safe Transportation Lab (*Zheng et al., 2019*). For example, information on vehicle position (i.e., coordinates of centroid), dimensions (length and width), orientation, average speed, and acceleration rate can be obtained. For the calculation of TTC, readers may refer to the formulations given in **Appendix**.

After the manual inspection and verification by experienced surveyors, trajectories of 2,217 vehicles were extracted. **Table 1** summaries the distribution of the sample, with respect to vehicle class and toll payment type. In particular, vehicles are classified into five categories: (i) private car, (ii) taxi, (iii) goods vehicle[1], (iv) bus, and (v) motorcycle. In addition, about half of the sample are using electronic toll payment (Count: 1,132; Proportion: 51.1%). **Figures 4** and **5** illustrate the vehicle trajectories for different toll payment types and vehicle classes, respectively.

---

[1] In this study, goods vehicle refers to light goods vehicle (excluding van-type vehicle), medium goods vehicle, and heavy goods vehicle.

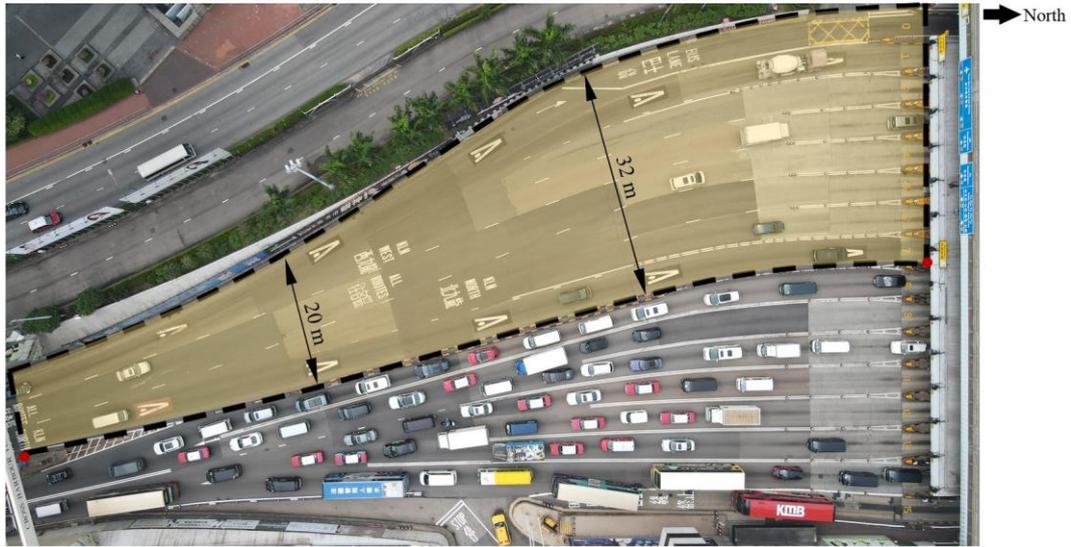

(a) Aerial view

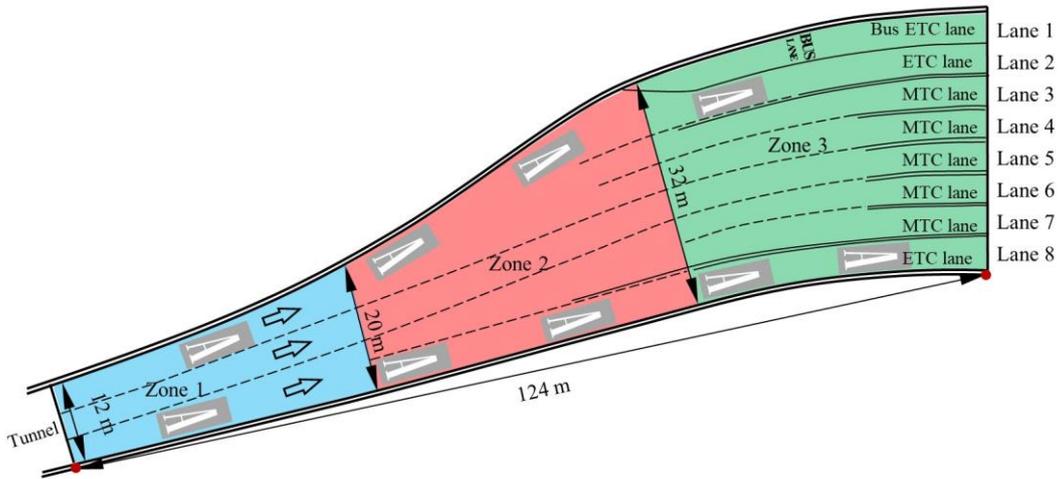

(b) Layout plan

**Figure 3 Layout of study site**

**Table 1 Distributions of the sample by toll payment type and vehicle class**

| Vehicle class | Toll payment type | | Overall |
|---|---|---|---|
| | Manual | Electronic | |
| Private car | 709 (32.0%) | 835 (37.6%) | **1544 (69.6%)** |
| Taxi | 175 (7.9%) | 20 (0.9%) | **195 (8.8%)** |
| Goods vehicle | 109 (4.9%) | 101 (4.6%) | **210 (9.5%)** |
| Bus | 5 (0.2%) | 164 (7.4%) | **169 (7.6%)** |
| Motorcycle | 87 (3.9%) | 12 (0.5%) | **99 (4.5%)** |
| **Total** | **1085 (48.9%)** | **1132 (51.1%)** | **2217 (100.0%)** |

It should be noted that the endogeneity issue is often overlooked in conflict analysis (*Yuan et al., 2022a*; *Yuan et al., 2022b*). There has been trade-off between prediction and casusality for the inclusion of an endogenous variable in model estimation (*Mannering et al., 2020*). To capture as many explanatory variables that are recognized to affect safety (*Mannering and Bhat, 2014*), it is crucial to use speed-related variables in conflict analysis. Hence, optimal traffic management and control measures can be implemented to mitigate the real-time crash risk (*Formosa et al., 2020*; *Mohammadian et al., 2021*; *Fu and Sayed, 2022*). To address the endogeneity issue, while the instantaneous speed of leading and following vehicles are used to calculate time-to-collision, average speed in preceding one second (i.e., 30 frames) of conflict vehicles are used as explanatory variables in the model. **Table 2** shows the descriptive statistics of variables considered. Angular speed refers to the rate of change in vehicle direction (degree per second), where clockwise is considered as positive.

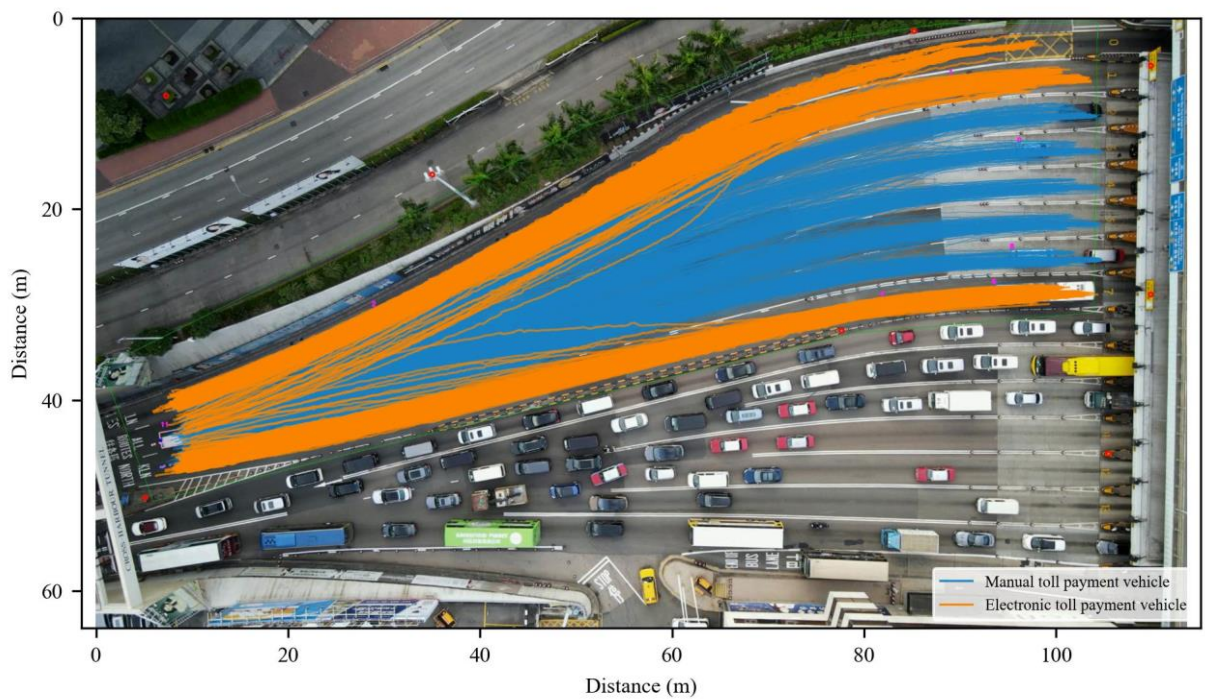

**Figure 4 Vehicle trajectories for different toll payment types**

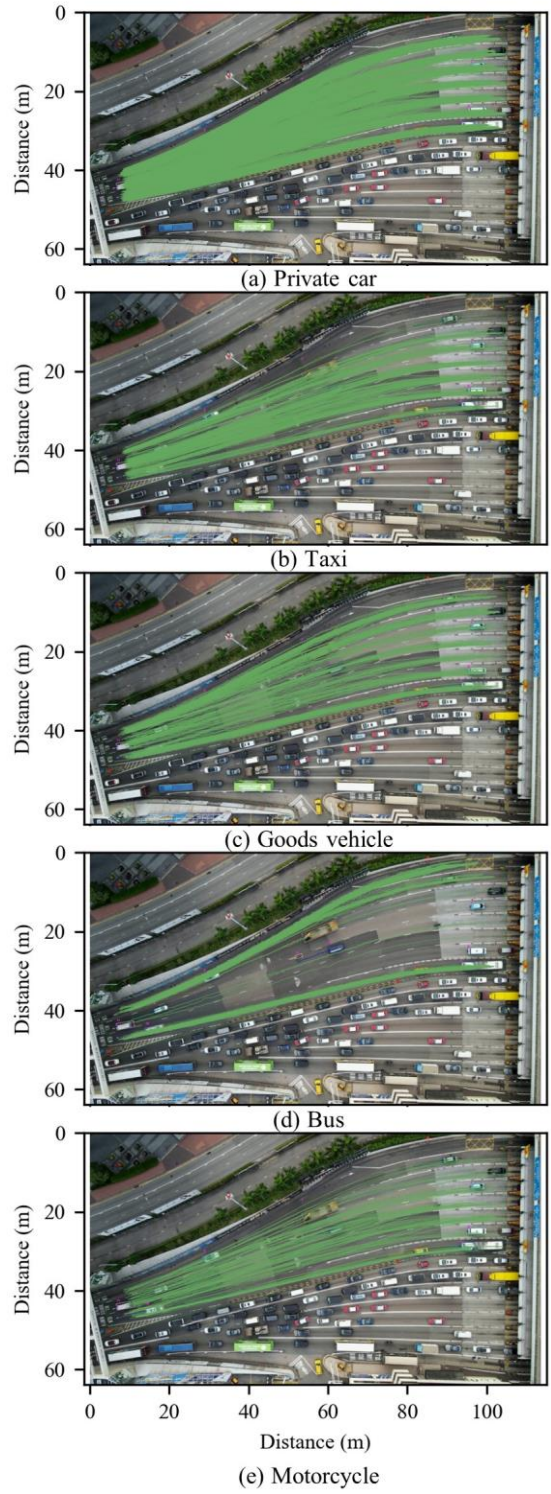

**Figure 5 Vehicle trajectories for different vehicle classes**

Table 2 Descriptive statistics of variables considered

| Factor | | Rear-end interaction and conflict | | | | Sideswipe interaction and conflict | | | |
|---|---|---|---|---|---|---|---|---|---|
| | | Mean | S.D. | Min. | Max. | Mean | S.D. | Min. | Max. |
| At least one vehicle uses electronic toll payment = 1; Otherwise = 0 | | 0.28 | 0.45 | 0 | 1 | 0.44 | 0.50 | 0 | 1 |
| Zone 1 | | 0.39 | 0.49 | 0 | 1 | 0.37 | 0.48 | 0 | 1 |
| Zone 2 | | 0.33 | 0.47 | 0 | 1 | 0.40 | 0.49 | 0 | 1 |
| Zone 3 | | 0.28 | 0.45 | 0 | 1 | 0.23 | 0.42 | 0 | 1 |
| Leading vehicle | Average speed (meter/second) | 8.56 | 3.17 | 0.59 | 21.6 | 11.01 | 2.79 | 2.46 | 19.54 |
| | Acceleration (meter/second$^2$) | -0.82 | 1.04 | -15.96 | 3.69 | -0.44 | 2.01 | -19.95 | 24.93 |
| | Angular speed (degree/second) | 4.46 | 2.55 | 0 | 34.18 | 4.75 | 3.07 | 0.0 | 28.36 |
| | Private car | 0.66 | 0.47 | 0 | 1 | 0.68 | 0.47 | 0 | 1 |
| | Taxi | 0.09 | 0.29 | 0 | 1 | 0.08 | 0.27 | 0 | 1 |
| | Goods vehicle | 0.19 | 0.39 | 0 | 1 | 0.13 | 0.34 | 0 | 1 |
| | Bus | 0.03 | 0.17 | 0 | 1 | 0.05 | 0.22 | 0 | 1 |
| | Motorcycle | 0.03 | 0.17 | 0 | 1 | 0.06 | 0.23 | 0 | 1 |
| Following vehicle | Average speed (meter/second) | 12.41 | 2.60 | 2.53 | 20.86 | 11.81 | 2.75 | 1.41 | 24.47 |
| | Acceleration (meter/second$^2$) | -0.26 | 3.10 | -19.55 | 48.12 | -0.24 | 3.88 | -87.75 | 37.15 |
| | Angular speed (degree/second) | 4.32 | 2.96 | 0 | 28.82 | 5.04 | 3.61 | 0.0 | 28.27 |
| | Private car | 0.63 | 0.48 | 0 | 1 | 0.66 | 0.48 | 0 | 1 |
| | Taxi | 0.13 | 0.34 | 0 | 1 | 0.08 | 0.27 | 0 | 1 |
| | Goods vehicle | 0.09 | 0.29 | 0 | 1 | 0.09 | 0.28 | 0 | 1 |
| | Bus | 0.02 | 0.13 | 0 | 1 | 0.02 | 0.14 | 0 | 1 |
| | Motorcycle | 0.13 | 0.33 | 0 | 1 | 0.16 | 0.36 | 0 | 1 |
| Number of observations | No conflict | 2726 (77.1%) | | | | 347 (24.5%) | | | |
| | Slight conflicts | 698 (19.7%) | | | | 692 (48.8%) | | | |
| | Severe conflicts | 111 (3.1%) | | | | 378 (26.7%) | | | |
| | Total observations | 3535 | | | | 1417 | | | |

## 4. ESTIMATION RESULTS AND DISCUSSION

Correlated grouped random parameter models with heterogeneity in the means are established for rear-end and sideswipe conflicts respectively. **Table 3** shows the results of goodness-of-fit of correlated and uncorrelated grouped random parameter multinomial logit models with heterogeneity in the means for rear-end and sideswipe conflicts. Results of likelihood ratio test indicate that correlated grouped random parameter multinomial logit models (Chi-square test statistic is 25.22 for rear-end conflict and 10.76 for sideswipe conflict) are superior, compared to the uncorrelated models, both at the 1% level of significance.

Table 3 Model performance metrics between uncorrelated and correlated models

| Metric | Rear-end conflicts model | | Sideswipe conflicts model | |
| --- | --- | --- | --- | --- |
| | Uncorrelated model | Correlated model | Uncorrelated model | Correlated model |
| McFadden $R^2$ | 0.59 | 0.59 | 0.25 | 0.26 |
| Degrees of freedom | 26 | 27 | 21 | 22 |
| Log likelihood | -1590.06 | -1577.45 | -1162.69 | -1157.31 |
| AIC | 3232.1 | 3210.9 | 2367 | 2359 |
| Chi-square test statistic | 25.22* | | 10.76* | |

Note: * Statistical significance at the 1% level

As shown in **Table 4**, likelihoods of rear-end conflicts (slight conflict: $\beta = 2.98$, severe conflict: $\beta = 1.89$) significantly increase when at least one electronic toll payment vehicle is involved. This could be because users of electronic toll payment tend to be more determined for the lane choice, and they may drive aggressively (they do not have to stop at the toll booths). As also shown in **Figure 3**, electronic toll collection lanes are located at the two sides of toll plaza (i.e., Lane 1, Lane 2, and Lane 8). Frequent lane changes may be required for the users of electronic toll payment (to reach the toll booths). On the other hand, users of manual toll payment may drive cautiously as they must keep checking the queue length of different toll booths, looking for that with the shortest queue (*Hong Kong Transport Department, 2020*). In addition, subscription fee is required for electronic toll payment in Hong Kong. Drivers using electronic toll payment are also frequent users of the tunnel. Hence, they should be more familiar with the route and traffic environment, compared to users of manual toll payment. Previous studies indicate that driver attention would decrease when route familiarity increases. Also, risk-taking behavior would be prevalent (*Martens and Fox, 2007*; *Charlton and Starkey, 2013*; *Weng et al., 2014*; *Intini et al., 2018*).

For the effects of vehicle dynamics, likelihood of rear-end conflict (slight conflict: $\beta = -0.40$; severe conflict: $\beta = -0.67$) significantly decreases when the average speed of leading vehicle increases. In contrast, average speed of following vehicle is positively associated with the likelihood of rear-end conflict (slight conflict: $\beta = 0.22$; severe conflict: $\beta = 0.42$). This should align with the failure mechanism of traffic conflicts and crashes (*Xing et al., 2020a*; *Arun et al., 2021*; *Zheng et al., 2021*). Just, it is important to recognize the potential endogeneity issue in real-time conflict analysis. In particular, one should be cautious about the model interpretation when speed-related variables, which were often used to calculate time-to-collision, are included as the explanatory variables in the model (*Mannering and Bhat, 2014*; *Mannering et al., 2020*). In addition, effects of average speed on conflict risk are normally distributed. For example, as shown in **Table 5**, there is a 26.6% chance that likelihood of severe sideswipe conflict would increase when the average speed of leading vehicle increases. This is because vehicles involving in sideswipe conflicts are not necessarily travelling on the same traffic lane (*Jiménez et al., 2013*).

Furthermore, effects of acceleration rates on the likelihood of rear-end conflicts are captured. For instance, the likelihood of slight rear-end conflict is negatively associated with the acceleration rate of leading vehicle, while the risk of severe rear-end conflict would increase when the acceleration rate of following vehicle increases. This could be attributed to the geometric design and traffic characteristics of toll plaza area. As the toll plaza is close to the portal of underwater tunnel portal, and the road segment (a diverging area with frequent lane changing and weaving activities) connecting tunnel portal and toll booths is a moderate crest curve, sight distance can be constrained. Therefore, positive association between accelerate rate of following vehicle and conflict risk is prevalent. Such finding is indicative to the implementation of remedial measures like advanced warning signs, pavement markings, and rumble strips that can increase the awareness of driver and overall safety (*Wong et al., 2006*; *Wong et al., 2012*). Effects of angular speeds on the likelihood of rear-end conflicts also show unobserved heterogeneity. For example, there is a 20.9% chance that the risk of severe rear-end conflict would increase when the angular speed of following vehicle increases. This could be attributed to the compensatory strategy adopted by the driver when one perceives that the lateral and steering controls are unstable (*Chen et al., 2021*), especially for the driver of following vehicle.

For the effect of vehicle class, likelihood of rear-end conflict increases when the following vehicle is a goods vehicle. Also, likelihood of sideswipe conflict increases when the following vehicle is a good vehicle. This could be attributed to the prevalence of blind spots and reduced vision for the drivers of heavy goods vehicles. It is difficult for the drivers to observe surrounding traffic, especially for weaving and lane-changing. Therefore, awareness and attentiveness of drivers could be impaired (*Cook et al., 2011*; *Marshall et al., 2020*). Such finding is indicative to vehicle design and innovations like advanced driver assistance system that could mitigate the risk attributed to reduced vision and inattentiveness (*Summerskill et al., 2016*). In addition, likelihood of sideswipe conflict of taxi is lower. This could be because of the compensatory behaviors adopted by the professional drivers who usually have better hazard perception skills (*Borowsky and Oron-Gilad, 2013*). Furthermore, risk of rear-end and sideswipe conflicts of motorcycle is higher. It could be because motorcyclists are more aggressive and risk-taking in general. As revealed in the crash statistics, crash involvement rates of motorcycle have been the highest among all vehicle classes in Hong Kong (*Hong Kong Transport Department, 2020*). Effective enforcement and educational strategies targeting to vulnerable driver groups, like commercial vehicle drivers and repeated conviction of traffic rules, can be implemented by the transport operators and government authorities. Therefore, safe driving can be promoted, and crash risk can be mitigated in the long run (*Chen et al., 2020*).

For the effect of vehicle's spatial location, likelihoods of rear-end and sideswipe conflicts at Zone 1 and Zone 2 are significantly lower, compared to Zone 3, at which lane changing activities are constrained. This may be contradictory to the findings of previous studies that crash risk is positively associated with lane changing and weaving activities at the merging areas (*Arbis and Dixit, 2019*; *Gu et al., 2019*), diverging areas (*Xing et al., 2020b*), and road work zones (*Park et al., 2018*; *Weng et al., 2018*; *Gan et al., 2020*). It could be because of the interactions between the vehicles stopped at Zone 3 to wait for the toll payment and those approaching from the tunnel portal. Speeds of the latter are usually higher. Also, driver capability to maintain lateral and steering stability could be impaired when the pavement markings are erased (*Chang et al., 2019*). Nevertheless, it is worth exploring on the association between geometric design, lane-changing activities, driver capability, and potential crash risk when information on visual perception and visual motor skills is available in naturalistic driving study and driving simulator experiment (*Chen et al., 2021*).

Furthermore, heterogeneity in the means of the random parameters is also considered. Factors that affect the means of the random parameters are identified. For example, for the likelihood of severe rear-end conflict, mean of the random parameter of angular speed is lower when an interaction is at Zone 1. For the likelihood of slight rear-end conflict, even that the mean coefficient of "following vehicle being a goods vehicle" is not significant ($\boldsymbol{\beta}_j$ being zero), heterogeneity in the mean $\boldsymbol{\theta}_j$ is statistically significant (*Alnawmasi and Mannering, 2019*). Mean of the random parameter of "following vehicle being a goods vehicle" increases when an interaction is at Zone 1. For the likelihood of slight and severe sideswipe conflict, means of the random parameter of "average speed of following vehicle" increase when an interaction is at Zone 2. Heterogeneity in the mean estimation indicate the possible correlation between random parameter and exogenous variables (*Mannering et al., 2016*). In this study, intervention effect by spatial location on the association between random parameter of vehicle dynamics and likelihood of conflict is revealed. Results are indicative to the real-time estimation of traffic conflict risk (*Alsaleh and Sayed, 2021*).

Correlations between random parameters are also considered. **Table 6-9** present the estimates of Cholesky matrix and correlation coefficient matrix. For the rear-end conflicts, as shown in **Table 6**, there is no significant correlation between the random parameters for slight rear-end conflict ("average speed of leading vehicle" and "following vehicle being a goods vehicle"), off-diagonal elements for slight rear-end conflict are constrained to zero. As shown in **Table 7**, for the likelihood of severe rear-end conflict, random parameter of average speed of leading vehicle is negatively correlated to that of the angular speed of following vehicle ($\gamma = -0.672$, $Cor = -0.96$). This implies the diminishing effects for the unobserved heterogeneity of average speed of leading vehicle on that of the angular speed of following vehicle. For the sideswipe conflicts, as shown in **Table 8**, off-diagonal elements of Cholesky matrix are not constrained. As shown in **Table 9**, for the likelihood of severe sideswipe conflict, there is negative correlation between the random parameters of average speeds of leading and following vehicles ($\gamma = -0.305$, $Cor = -0.77$). Again, effects of unobserved heterogeneity of these two factors on severe sideswipe conflict are offsetting. Such findings may be attributed to the compensatory strategies adopted by the drivers under emergency, especially for experienced drivers (*Chen et al., 2021*). Yet, it is worth exploring the effectiveness of advanced driver assistance system in improving the driver performance and mitigating the safety risk, especially for lane changing and weaving activities in diverging area, when comprehensive data on driver visual perception, and perception-motor skills are available from naturalistic driving study and driving simulator experiment (*Chen et al., 2019a*; *Chen et al., 2019b*; *Mohammadian et al., 2021*).

**Table 4 Results of parameter estimation of correlated model with heterogeneity in the means for rear-end conflicts**

|  | Slight conflict | | Severe conflict | |
| --- | --- | --- | --- | --- |
|  | Coefficient | t-stat | Coefficient | t-stat |
| Constant | -1.73 | -2.92 | -7.19 | -5.30 |
| At least one electronic toll payment vehicle involved | 2.98 | 6.73 | 1.89 | 2.19 |
| Zone 1 | -3.91 | -10.22 | -4.92 | -4.39 |
| Zone 2 | -1.44 | -6.07 | -3.31 | -4.57 |
| **Characteristics of leading vehicle** | | | | |
| Average speed | -0.40 | -5.79 | -0.94 | -3.38 |
| *Standard deviation* | 0.45 | 11.45 | 0.78 | 4.37 |
| Acceleration | -0.30 | -3.82 | - | - |
| Angular speed | -0.09 | -2.76 | - | - |
| Motorcycle | - | - | 4.57 | 4.55 |
| **Characteristics of following vehicle** | | | | |
| Average speed | 0.22 | 3.63 | 0.42 | 2.64 |
| Acceleration | - | - | 0.11 | 2.64 |
| Angular speed | - | - | -0.57 | -1.77 |
| *Standard deviation* | - | - | 0.70 | 2.78 |
| Goods vehicle | IS | IS | 2.11 | 2.22 |
| *Standard deviation* | 3.13 | 8.25 | - | - |
| Motorcycle | 1.74 | 5.50 | 3.19 | 4.84 |
| **Heterogeneity in the means of the random parameter** | | | | |
| Following vehicle being a goods vehicle: Zone 1 | 2.94 | 3.27 | - | - |
| Angular speed of following vehicle: Zone 1 | - | - | -0.59 | -3.68 |
| **Model statistics** | | | | |
| McFadden R² | 0.59 | | | |
| Number of observations | 3535 | | | |
| Degree of freedom | 27 | | | |
| Log-likelihood at zero (LL(0)) | -3883.59 | | | |
| Log-likelihood at convergence (LL(β)) | -1577.45 | | | |

Notes: IS: not significant.

**Table 5 Results of parameter estimation of correlated model with heterogeneity in the means for sideswipe conflicts**

| Variable | Slight conflict | | Severe conflict | |
|---|---|---|---|---|
| | Coefficient | t-stat | Coefficient | t-stat |
| Zone 1 | -1.91 | -5.15 | -3.94 | -6.74 |
| Zone 2 | -2.77 | -2.82 | -4.10 | -2.62 |
| **Characteristics of leading vehicle** | | | | |
| Average speed | - | - | -0.25 | -2.86 |
| *Standard deviation* | - | - | 0.40 | 4.42 |
| Bus | - | - | 3.23 | 3.78 |
| Motorcycle | 1.69 | 3.06 | 2.18 | 2.84 |
| **Characteristics of following vehicle** | | | | |
| Average speed | 0.17 | 2.76 | 0.30 | 2.11 |
| *Standard deviation* | 0.13 | 8.53 | 0.40 | 4.95 |
| Taxi | - | - | -1.92 | -2.59 |
| Goods vehicle | 1.10 | 2.60 | 1.36 | 1.87 |
| Motorcycle | 0.90 | 1.93 | - | - |
| **Heterogeneity in the means of the random parameter** | | | | |
| Average speed of following vehicle: Zone 2 | 0.20 | 2.17 | 0.25 | 1.93 |
| **Model statistics** | | | | |
| McFadden R² | 0.26 | | | |
| Number of observations | 1417 | | | |
| Degree of freedom | 22 | | | |
| Log-likelihood at zero (LL(0)) | -1556.73 | | | |
| Log-likelihood at convergence (LL(β)) | -1157.31 | | | |

**Table 6 Cholesky matrix of random parameters for rear-end conflict (t-statistic in parentheses)**

| Variable | Severe conflict | |
|---|---|---|
| | Average speed of | Angular speed of |

|  |  | leading vehicle | following vehicle |
|---|---|---|---|
| Severe conflict | Average speed of leading vehicle | 0.783 (4.37) | 0 |
|  | Angular speed of following vehicle | -0.672 (-2.58) | 0.207 (4.50) |

**Table 7 Correlation coefficient matrix of random parameters for rear-end conflict**

| Variable |  | Severe conflict ||
|---|---|---|---|
|  |  | Average speed of leading vehicle | Angular speed of following vehicle |
| Severe conflict | Average speed of leading vehicle | 1.00 | -0.96 |
|  | Angular speed of following vehicle | -0.96 | 1.00 |

**Table 8 Cholesky matrix of random parameter for sideswipe conflict (t-statistic in parentheses)**

| Variable |  | Severe conflict ||
|---|---|---|---|
|  |  | Average speed of leading vehicle | Average speed of following vehicle |
| Severe conflict | Average speed of leading vehicle | 0.397 (4.42) | 0 |
|  | Average speed of following vehicle | -0.305 (-3.57) | 0.254 (6.12) |

**Table 9 Correlation coefficient matrix of random parameters for sideswipe conflict**

| Variable |  | Severe conflict ||
|---|---|---|---|
|  |  | Average speed of leading vehicle | Average speed of following vehicle |
| Severe conflict | Average speed of leading vehicle | 1.00 | -0.77 |
|  | Average speed of following vehicle | -0.77 | 1.00 |

## 5. CONCLUSIONS

This study examines the safety risk of tunnel toll plaza based on the high-resolution trajectory data captured using drone. The correlated grouped random parameters multinomial logit model with heterogeneity in the means is adopted, accounting for the effects of repeated observations, unobserved heterogeneity, and correlation among random parameters at the road user level. Associations between possible influencing factors, occurrence and severity of traffic conflicts at the tunnel toll plaza are measured. In particular, modified traffic conflict indicator is proposed to account for the effects of dimensions (both width and length) and longitudinal and angular movement of interacting vehicles when estimating the conflict risk. This should improve the accuracy of conflict risk estimation, compared to the conventional (vehicle) centroid-based approach. In addition, effect of traffic conflict type (rear-end and sideswipe conflicts) on the association is considered.

Results indicate that when at least one electronic toll payment user is involved, likelihood of slight rear-end conflict would increase, but that of severe rear-end conflict would decrease. As expected, likelihood of conflict is negatively associated with the average speed of leading vehicle, and positively associated with that of following vehicle. However, effects of average speed, acceleration rate, and angular speed on the conflict risks are random. These could be attributed to the compensatory behavior adopted by the drivers in emergency. Furthermore, conflict risks generally increase when goods vehicle and motorcycle are involved. This may be because of the reduced vision of goods vehicle drivers and risk-taking behaviors of motorcyclists. Nevertheless, correlated approach with heterogeneity in the means allows additional flexibility when capturing unobserved heterogeneity at the road user level. There are negative correlations between the random parameters of severe rear-end and sideswipe conflicts.

It is recommended that vehicle design could be enhanced, and driver assistance system could be introduced to mitigate the risk attributed to the reduced vision of drivers, especially for heavy vehicles including buses and heavy goods vehicles. Findings are also indicative to the remedial design and measures for tunnel toll plazas including lane markings and advanced warning signs that can guide the drivers to the correct toll booths, and therefore reduce the risks of conflicts attributed to frequent lane changing and weaving activities (*Wong et al., 2006*). Nevertheless, it is worth investigating for the effects of geometric design, and configurations (i.e., lane allocation, traffic signs, pavement markings, and speed limit) on the risk of conflict when the trajectory data at other locations are available. Furthermore, effects of innovations like advanced driver assistance system on driver attention, visual perception, and perception-motor skill could be explored when comprehensive driving data are obtained from naturalistic driving study and driving simulator experiment in the future. Last but not least, it is crucial to address the endogeneity issue in the model using appropriate statistical corrections (*Guevara and Ben-Akiva, 2012*).

# ACKNOWLEDGEMENTS

The work that is described in this paper as supported by Research Committee of the Hong Kong Polytechnic University (H-ZJMQ) and the Smart Traffic Fund (PSRI/09/2108/PR). We would like to thank the Smart and Safe Transportation Lab of University of Central Florida (UCF-SST) that have helped us to improve the quality of the paper, particularly vehicle trajectory detection and tracking.

# APPENDIX

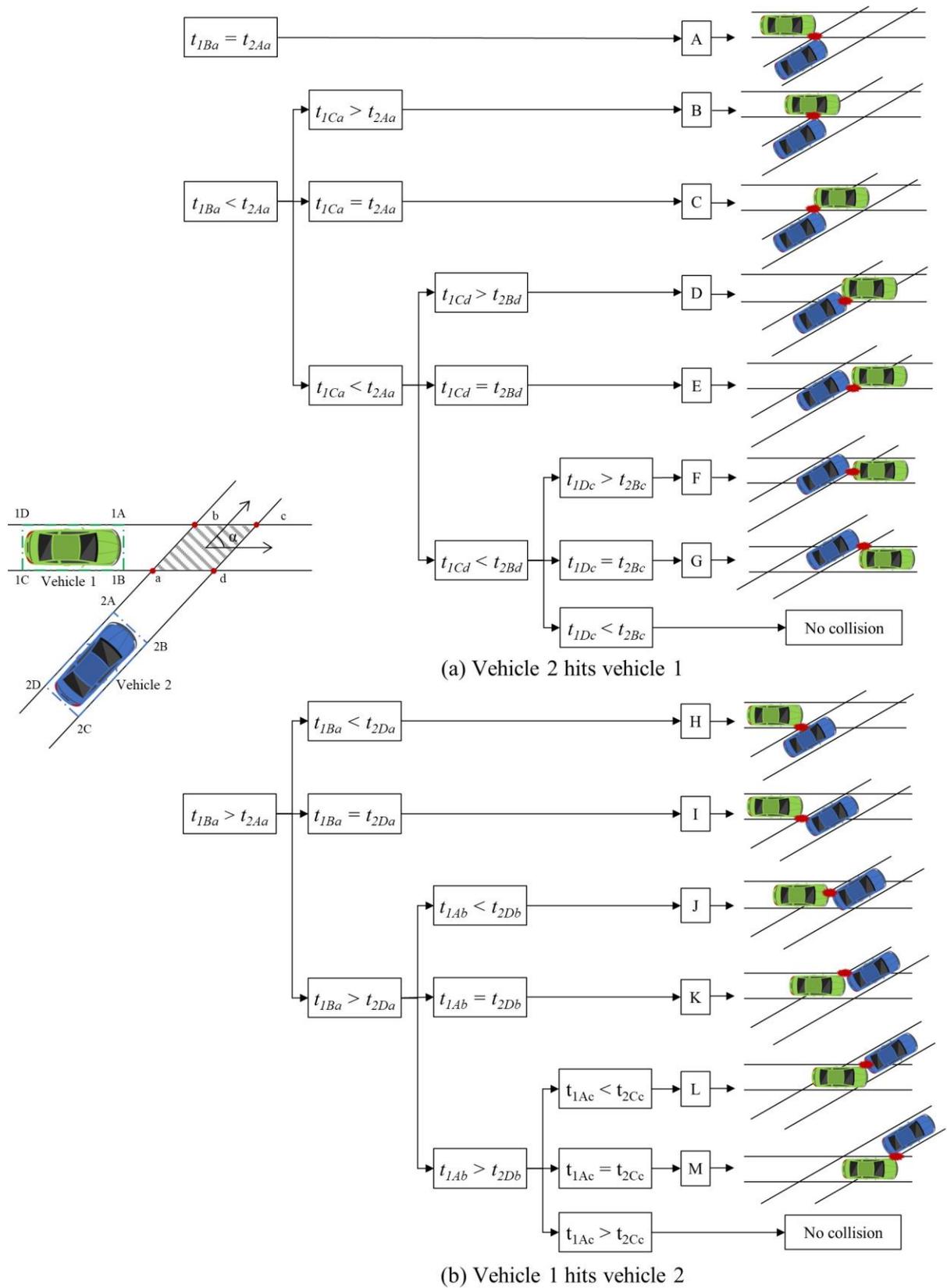

Figure A1 Illustration of possible conflict scenarios

```
if t_{1BA} < t_{2Aa} then
    if t_{1Ca} < t_{2Aa} then
        if t_{1Cd} < t_{2Bd} then
            if t_{1Dc} > t_{2Bc} then
```
$$TTC = \frac{t_{2Bd} \times t_{1Dc} - t_{1Cd} \times t_{2Bc}}{t_{2Bd} + t_{1Dc} - t_{1Cd} - t_{2Bc}}$$
```
            else if t_{1Dc} = t_{2Bc} then
                TTC = t_{2Bc} (or t_{1Dc})
        else if t_{1Cd} > t_{2Bd} then
```
$$TTC = \frac{t_{1Cd} \times t_{2Aa} - t_{2Bd} \times t_{1Ca}}{t_{1Cd} + t_{2Aa} - t_{2Bd} - t_{1Ca}}$$
```
        else
            TTC = t_{2Bd} (or t_{1Cd})
    else if t_{1Ca} > t_{2Aa} then
        TTC = t_{2Aa}
    else
        TTC = t_{2Aa} (or t_{1Ca})
else if t_{1BA} > t_{2Aa} then
    if t_{1Ba} > t_{2Da} then
        if t_{1Ab} > t_{2Db} then
            if t_{1Ac} < t_{2Cc} then
```
$$TTC = \frac{t_{2Cc} \times t_{1Ab} - t_{1Ac} \times t_{2Db}}{t_{2Cc} + t_{1Ab} - t_{1Ac} - t_{2Db}}$$
```
            else if t_{1Ac} = t_{2Cc} then
                TTC = t_{1Ac} (or t_{2Cc})
        else if t_{1Ab} < t_{2Db} then
```
$$TTC = \frac{t_{2Db} \times t_{1Ba} - t_{1Ab} \times t_{2Da}}{t_{2Db} + t_{1Ba} - t_{1Ab} - t_{2Da}}$$
```
        else
            TTC = t_{1Ab} (or t_{2Db})
    else if t_{1Ba} < t_{2Da} then
        TTC = t_{1Ba}
    else
        TTC = t_{1Ba} (or t_{2Da})
else
    TTC = t_{1BA} (or t_{2Aa})
```

**Figure A2 Formulation for modified TTC**